\documentstyle[12pt]{article}

\begin{document}
\sloppy
\begin{flushright}{TIT/HEP-390\\ April '98}\end{flushright}
\vskip 1.5 truecm
\centerline{\large{\bf On the cosmological domain wall problem }}
\centerline{\large{\bf in supersymmetric models}}
\vskip .75 truecm
\centerline{\bf Tomohiro Matsuda
\footnote{matsuda@th.phys.titech.ac.jp}}
\vskip .4 truecm
\centerline {\it Department of physics, Tokyo institute of technology}
\centerline {\it Oh-okayama, Meguro,Tokyo 152-8551, Japan}
\vskip 1. truecm
\makeatletter
\@addtoreset{equation}{section}
\def\theequation{\thesection.\arabic{equation}}
\makeatother
\vskip 1. truecm
\begin{abstract}
\hspace*{\parindent}
Many models for dynamical supersymmetry breaking have the domain wall 
solution of the Kovner-Shifman type and therefore contain the potential
cosmological domain wall problem.
We show that the degeneracy of the vacua is lifted in any model
in which the cosmological constant is fine-tuned to zero through
an explicit $Z_{n}^{R}$ symmetry breaking constant in the superpotential.
In this respect, we do not have to add unnatural non-renormalizable
terms which explicitly break the symmetry. 
We also discuss the conditions for the safe decay of the walls
for several concrete examples.

\end{abstract}
\newpage
\section{Introduction}
\hspace*{\parindent}
In general many models for supersymmetry breaking have  domain wall
solutions of the Kovner-Shifman type\cite{Kovner_Shifman} in
their dynamical sectors where the discrete symmetry is
implemented by their anomalies.
A domain wall configuration may also appear as the 
consequence of spontaneous symmetry breaking of the explicit
$Z_{n}^{R}$ symmetry
which  is imposed by hand in order to solve the 
phenomenological difficulties such as the $\mu$-problem\cite{Abel} or
the cosmological moduli problem\cite{Asaka}.
These domain walls are generally dangerous from the cosmological 
point of view.
If the Universe undergoes a phase transition associated with the
spontaneous symmetry breaking, domain walls will inevitably form
and in most cases these domain walls are very dangerous for
the standard evolution of the Universe.

In this paper we study the properties of the Kovner-Shifman 
domain walls in general supersymmetric  models.
We show that the degeneracy of the vacua is always lifted by a constant
term when supergravity is turned on.
The magnitude of the energy difference is estimated to be 
$\sim \sigma^{2}/M_{p}^{2}$
where $\sigma$ is the surface energy density of the wall.
This value is sufficient for the safe decay of the cosmological 
domain wall provided $\sigma$
is larger than $(10^{5}GeV)^{3}$.
Although the model is technically non-generic because it includes
a single term which explicitly breaks the $Z_{n}^{R}$ symmetry, 
it is still a reasonable model for a vanishing cosmological constant.

The basic idea is very similar to the well-known mechanism for 
the mass generation of the R-axion\cite{Axion}

\section{Kovner-Shifman Domain walls}
\hspace*{\parindent}
In this section we explain the basic idea of the Kovner-Shifman domain
wall\cite{Kovner_Shifman}.
The Kovner-Shifman domain wall is a configuration which preserves
a half of the supersymmetry. 
The key ingredient is the central extension of the N=1 superalgebra
\begin{eqnarray}
\label{algebra}
\{\overline{Q}_{\dot{\alpha}},\overline{Q}_{\dot{\beta}}\}&=&
        -2i(\sigma^{0})^{\gamma}_{\dot{\alpha}}\int d^{3}x
        \{\overline{D}_{\dot{\beta}}\overline{D}^{\dot{\delta}}
        J_{\gamma\dot{\delta}}\}_{\theta=0}.
\end{eqnarray}
The contribution to the central extension is due to the non-conservation
of the  current  multiplet $J$.
The lowest component of $J$ is the $U(1)_{R}$-current which is
broken classically (in Wess-Zumino model with more than two terms in 
the superpotential)
or by quantum anomaly ($U(1)_{R}$ anomaly in gauge theories).
Although the current multiplet $J$ have anomaly, the supersymmetry 
current is conserved thus supersymmetry is still unbroken.
What is not conserved in this case
is not the supersymmetry current but the superconformal current.

\underline{ Domain wall in non-gauge theories}

The simplest example for the Kovner-Shifman domain wall is the  
Wess-Zumino model with only a superfield $\Phi$.
In non-gauge theories, the presence of the central extension
can be seen at the tree level.
The Wess-Zumino Lagrangian in terms of a superfield $\Phi$
has the following form,
\begin{eqnarray}
L&=&\frac{1}{4}\int d^{4}\theta \Phi \overline{\Phi}+\left[
        \frac{1}{2}\int d^{2}\theta W(\Phi)+h.c.\right],\nonumber\\
W(\Phi)&=&\mu^{2}\Phi-\frac{\lambda}{3}\Phi^{3}.
\end{eqnarray}
When $\mu=0$, the Lagrangian is invariant under the $U(1)_{R}$ rotation
\begin{eqnarray}
  d\theta&\rightarrow& e^{-i\alpha}d\theta,\nonumber\\
  \Phi&\rightarrow&  e^{\frac{2i\alpha}{3}}\Phi.
\end{eqnarray}
However, if $\mu\ne0$, this $U(1)_{R}$ invariance is broken and
only the discrete part $Z_{2}^{R}$ persists,
\begin{eqnarray}
  d^{2}\theta&\rightarrow& -d^{2}\theta,\nonumber\\
  \Phi&\rightarrow&  -\Phi.
\end{eqnarray}
When $<\Phi>\ne 0$, the $Z_{2}^{R}$ symmetry is spontaneously broken
and the corresponding
domain wall which interpolates between two degenerate vacua 
\begin{eqnarray}
\label{wz}
<\Phi>&=&\pm \mu/\sqrt{\lambda},\nonumber\\
<W>&=&\mp\frac{2}{3}\frac{\mu^{3}}{\sqrt{\lambda}}
\end{eqnarray}
appears.
In this case the supersymmetric current multiplet $J$
has the anomaly of the following form,
\begin{eqnarray}
\overline{D}^{\dot{\alpha}}J_{\alpha\dot{\alpha}}&=&
\frac{1}{3}D_{\alpha}\left[3W-\Phi\frac{\partial W}{\partial \Phi}\right].
\end{eqnarray}
Substituting the anomaly equation into the central extension of the 
superalgebra(\ref{algebra}), one obtains
\begin{eqnarray}
\{Q_{\alpha},Q_{\beta}\}&=&4(\vec{\sigma})_{\alpha\beta}
\int d^{3} x\vec{\nabla}\left[\overline{W}-\frac{1}{3}\overline{\Phi}
        \frac{\partial\overline{W}}{\partial\overline{\Phi}}\right]
        _{\overline{\theta}=0}
\end{eqnarray}
where the right hand side of the equation is related to the central
charge of the wall configuration therefore it represents  the surface
energy density
of the wall configuration.
It is apparent that the contribution is non-zero when the domain wall
interpolates the two different vacuum configurations of eq.(\ref{wz}).

The domain wall solution for such non-gauge theories 
can easily be extended to more general
theories such as an effective theory of SQCD or some other complicated
theories.
For later convenience we show one of such examples which contains
explicit cut-off parameter of the theory and have the superpotential with
discrete $Z_{n}^{R}$ symmetry,
\begin{eqnarray}
d^{2}\theta&\rightarrow&d^{2}\theta e^{-\frac{2a\pi i}{n}}\nonumber\\
X&\rightarrow& X e^{\frac{2\pi i}{n}}.
\end{eqnarray}
The superpotential takes a form
\begin{equation}
\label{ZnR}
W=\sum^{\infty}_{k=0}\frac{\lambda_{k}}{nk+a}{X^{nk+a}}{
        M^{-nk-a+3}},
\end{equation}
where $\lambda_{k}$ and $M$  denotes the 
coupling constants and the cut-off
scale of the model.
$a$ is an integer which will be fixed in the corresponding
phenomenological models.
If the coupling constants are assigned so that the scalar potential 
exhibits  non-trivial minima at $<X>\sim Me^{\frac{2\pi i}{n}k}
(k=1,..n$), the discrete symmetry
is spontaneously broken.
A domain wall solution which interpolates the degenerated
$Z_{n}^{R}$ vacuum configurations appears and
the wall have the surface energy density of order $M^{3}$.

\underline{Domain wall in gauge theories}

Now let us discuss the supersymmetric QCD(SQCD) with or without
the superpotential.
As we have stated above, this theory have an anomaly which contributes
to the central extension.
The Lagrangian is
\begin{eqnarray}
L&=&\left[\frac{1}{4g^{2}}\int d^{2} \theta Tr{\cal W}^{2}+h.c.\right]
+\frac{1}{4}\int d^{2}\theta \left[\overline{Q}^{i}e^{V}Q^{i}+
\overline{\tilde{Q}}^{i}e^{V}\tilde{Q}^{i}\right].
\end{eqnarray}
In superfield notation the anomaly in the supermultiplet of current is
\begin{eqnarray}
\overline{D}^{\dot{\alpha}}J_{\alpha\dot{\alpha}}&=&
\frac{1}{3}D_{\alpha}\left\{\left[3W-\sum_{i}Q_{i}\frac{\partial W}
{\partial Q_{i}}\right]\right.\nonumber\\
&&\left.\left[\frac{3T(G)-\sum_{i}T(R_{i})}{16\pi^{2}}
Tr{\cal W}^{2}+\frac{1}{8}\sum_{i}\gamma_{i}Z_{i}\overline(D)^{2}
(\overline{Q}_{i}e^{V}Q_{i})\right]\right\}
\end{eqnarray}
where in SQCD,  $T(G)=N_{c}$ and $T(R_{i})=1$ for each flavor.
Substituting the anomaly equation into the superalgebra(\ref{algebra})
and using the Konishi anomaly equation, one gets
\begin{eqnarray}
\label{sqcd_cent}
\{\overline{Q}_{\alpha},\overline{Q}_{\beta}\}
&=&4(\vec{\sigma})_{\dot{\alpha}\dot{\beta}}
\int d^{3} x\vec{\nabla}\left[W-\frac{T(G)-\sum_{i}T(R_{i})}{16\pi^{2}}
Tr{\cal W}^{2}\right]_{\theta=0}.
\end{eqnarray}
For pure SYM without matter field, it is easy to find that the domain 
wall have non-zero central charge.
The contribution to the central charge appears because
gaugino condensation has the phase.
To be more precise, the surface energy density takes a form 
$\Lambda^{3}|1-e^{\frac{2\pi i}{N_{c}}}|$
for nonchiral-nonchiral configurations or 
$\Lambda^{3}|0-e^{\frac{2\pi i}{N_{c}}}|$ for chiral-nonchiral 
configuration.
For $N_{f}<N_{c}$ SQCD the domain wall solution is formulated for
the effective superpotential which looks like a extended
Wess-Zumino model.
It is easy to understand from eq.(\ref{sqcd_cent})
that there is always a contribution
to the central charge as far as 
gaugino condensation is non-zero.
(Note that  the coefficient in front of 
gaugino condensation does not vanish for $N_{f}< N_{c}$.)
Although the contribution from anomaly vanishes for $N_{f}=N_{c}$
massless SQCD, the
central charge does not vanish for massive SQCD because there is a
contribution from  the mass term \cite{matsuda}.

\section{Cosmological problem of the Kovner-Shifman domain wall}
\hspace*{\parindent}
In this section we discuss the cosmological problem of 
Kovner-Shifman domain walls and find their solutions.

It is well known  that whenever the Universe undergoes a phase
transition associated with the spontaneous symmetry breaking,
domain walls will inevitably form.
In most cases the domain walls are dangerous for the standard evolution
of the universe.

As we have discussed in the previous section, the  models for dynamical 
supersymmetry breaking or the models with explicit $Z_{n}^{R}$ 
symmetry  should have 
the Kovner-Shifman domain wall solution.

\underline{ Hidden sector models}

As the first example, let us consider the hidden sector models
in which the supersymmetry breaking is triggered by gaugino 
condensation in the hidden sector.
The typical scale of the surface energy density of the domain wall
is related to  gaugino
condensation in the hidden sector 
and is estimated to be $\sim 10^{12}$GeV which
is high enough to avoid the production of the cosmological domain wall
during the reheating of the Universe.

\underline{ Gauge mediated supersymmetry breaking models}

The second example is the gauge mediated supersymmetry breaking
model\cite{dsb}.
Although this type of models have so many variants, most of 
the models are inspired by the dynamical analysis of 
SQCD with $N_{f}\le N_{c}$ fundamental fields.
In this respect, here we may focus our attention only to SQCD with
$N_{f}\le N_{c}$ without loss of generality.
Contrary to the hidden sector models, the surface energy density of the
domain wall in the dynamical sector 
is very low.
The walls can easily be produced 
during the reheating of the Universe thus needs some mechanism
which makes the wall collapse in the proper time scale.
The pressure which  is required for the safe decay of
the wall must be induced by 
the explicit breaking of the degeneracy of the vacuum energy.
In the crudest estimate, the required value for the pressure is 
\cite{vilenkin}
\begin{equation}
\label{decay}
\epsilon\geq\frac{\sigma^{2}}{M_{p}^{2}}
\end{equation}
where $\epsilon$ represents the pressure corresponding to the difference
of the energy density between the corresponding vacua.
$\sigma$ represents
the surface energy density of the domain wall.

The most natural and simplest candidate for the source of $\epsilon$
is the constant term which arises from the general requirement
that the cosmological constant
must be canceled in any successful theory of dynamical supersymmetry
breaking.
In supergravity theories, at the tree level the scalar potential takes the
following form
\begin{eqnarray}
V&=&V_{D}+V_{F}\nonumber\\
V_{D}&=&\frac{1}{2}g^{2}D^{a}D^{a} \nonumber\\
V_{F}&=&e^{\frac{K}{M_{p}^{2}}}\left(\left[W_{i}+\frac{K_{i}}{M^{2}_{p}}
        W\right]K^{-1}_{ij^{*}}\left[W^{*}_{j^{*}}+\frac{K_{j^{*}}}
        {M^{2}_{p}}W^{*}\right]-3\frac{|W^{2}|}{M_{p}^{2}}\right),
\end{eqnarray}
where $W$ and $K$ are the superpotential and the K\"ahler potential.
In general models for supersymmetry breaking, a constant term $W_{0}$
is added and  is adjusted to cancel the vacuum energy.
Because we are assuming that $W_{0}$ is a constant, this fine tuning 
works only for one vacuum.
\footnote{To avoid the appearance of the negative energy state,
supersymmetry breaking parameter $F$ should satisfy the condition
$V=|F|^{2}-3|W_{tot}|^{2}/M_{p}^{2}\ge0$ for any k-th
  vacuum.
It is also required that the  cancellation  works for the 
lowest energy state of the k-th vacuum.)}
When the cancelation occurs for the $k_{0}$-th vacuum,
the situation is schematically given as:
\begin{eqnarray}
V_{True}&=&|F|^{2}-3\frac{|W_{0}+\Lambda^{3}e^{\frac{2\pi i}{n}
k_{0}}|}{M_{p}^{2}}\nonumber\\
&=&0\\
V_{False}&=&|F|^{2}-3\frac{|W_{0}+\Lambda^{3}e^{\frac{2\pi i}
{n}k}|}{M_{p}^{2}}\nonumber\\
&\ne&0 \hspace{2cm}\mbox{for $k\ne k_{0}$}\\
\epsilon&=&V_{False}-V_{True}\nonumber\\
&=&-3\frac{|W_{0}+\Lambda^{3}e^{\frac{2\pi i}{n}k}|}{M_{p}^{2}}+
3\frac{|W_{0}+\Lambda^{3}e^{\frac{2\pi i}{n}k_{0}}|}{M_{p}^{2}}
\nonumber\\
&\sim&\frac{\Lambda^{3}}{M_{p}^{2}}.
\end{eqnarray}
Here $\Lambda$ and $F$ denotes the  scale of the dynamical sector
and the supersymmetry breaking parameter.
(Note that the above equation
  gives the essential part of the scalar potential for our discussion
but is not the full description of the exact scalar
potential.)
Other vacua ($k\ne k_{0}$)
must have non-zero vacuum energy and it is estimated
to be of order $ \sim \sigma^{2}/M_{p}^{2}$.
In fact, this satisfies the required condition(\ref{decay})
for the safe decay of 
the cosmological domain wall.
In this case, the domain wall problem is solved without introducing 
an extra breaking term nor the fine-tuning of the parameter.

\underline{ Model for inflation}

The third example is the thermal inflation model of 
Asaka et al\cite{Asaka}.
This model is proposed by T.Asaka, J.Hashiba, M.Kawasaki and
T.Yanagida as a solution for the cosmological moduli problem
in gauge-mediated supersymmetry breaking theories.
In ref.\cite{Asaka}, $Z_{n+3}$ symmetry is imposed on the
superpotential for the flaton.
The vacua are degenerated and has $Z_{n+3}$ phase 
thus an explicit breaking of the discrete symmetry is needed
to remove the domain wall.
In ref.\cite{Asaka} a small term $\alpha X$ was added by hand to 
the superpotential so that the discrete symmetry is  explicitly broken.
However, as we have seen above, such an unnatural term is not 
necessitated if $Z^{R}_{n}$ is assigned instead of $Z_{n+3}$.
(See eq.(\ref{ZnR}).)

\underline{ Next-to-minimal supersymmetric Standard Model}

The last example is the next-to-minimal supersymmetric Standard
Model (NMSSM).
This model contains an additional singlet Higgs superfield
and $Z_{3}^{R}$ symmetry which transforms every chiral superfield
$\Phi$.
The $\mu$-term is generated because the singlet
that parametrizes the $\mu$-term 
gains  non-zero  vacuum expectation value.
In this model, the natural scale for the domain wall is the weak scale.
Although the energy gap is produced in the same way by
the constant term in the superpotential, one should consider
further constraint coming from primordial nucleosynthesis
because  the scale of the wall is smaller than
$10^{5}GeV$(see ref.\cite{Abel}).
In fact the weak-scale walls  decay after nucleosynthesis and the
entropy produced when the wall collide is dumped into all the decay 
products. 
In this case, because of this additional constraint,
the solution is not automatic.
One possibility is to raise the surface energy density
of the wall $\sigma$ above
$10^{5} GeV$.
However,  it requires fine-tuning of parameters  
because the $\mu$-term is already fixed at the weak scale.
Another possibility is to introduce a non-renormalizable
term that explicitly breaks the symmetry.
On the whole, unlike other walls discussed above,
the domain wall in NMSSM requires additional mechanisms or fine-tunings
of the parameters to evade the cosmological domain wall problem.

\underline{The origin of the constant term}

Although we did not mentioned about the origin of the constant term
$W_{0}$ in the above discussions, it would be fair to say
that such contribution  is very common.
One may think that the constant term appears from another dynamical
sector which have the potential $\Lambda^{3}e^{\frac{2\pi i}{n}}$.
The most obvious example is the simple SYM with the product gauge group
$SU(N_{1})\times SU(N_{2})$.
The resulting superpotential is expected to be of the form:
$W=\Lambda_{1}^{3}
e^{\frac{2\pi i}{N_{1}}k_{1}}+
\Lambda^{3}_{2}e^{\frac{2\pi i}{N_{2}}k_{2}}$, where
$\Lambda_{1,2}$ is the dynamical scale of each gauge sector.
Although we did not introduced an explicit constant term in this case,
the vacuum degeneracy is obviously lifted when supergravity is
turned on.

\section{Conclusion and discussions}
\hspace*{\parindent}
In general it is quite difficult to find  a successful model for 
supersymmetry breaking without introducing unnatural components.
In this paper we have shown that the existence of the domain wall
does not further constrain these models provided the surface energy
density
of the wall is not so low as the weak scale.
Our point is that the cosmological constant should be cancelled
by adding a constant term to the superpotential.
This constant explicitly breaks any continuous or discrete R-symmetry,
and lifts the degenerated vacua.

In this paper we have concerned with two types of walls that
is related to two kinds of   $Z_{n}^{R}$ symmetry.
One is the explicit symmetry of the Lagrangian, and the other
appears as the consequence of the breakdown of the continuous
$U(1)_{R}$ symmetry through anomaly.
The wall of the latter type is especially important because it
appears in almost all the dynamical supersymmetry breaking models.

\section{Acknowledgment}
We are very grateful to T.Asaka for many useful discussions.

\end{document}